\documentclass{article}
\usepackage{graphicx}%style pour inserer figures en eps
\usepackage{here}%pour inserer figures et tables a emplacement 
\def\beq{\begin{equation}}
\def\eeq{\end{equation}}
\def\bea{\begin{eqnarray}}
\def\eea{\end{eqnarray}}
\def\bq{\begin{quote}}
\def\eq{\end{quote}}

\def\nnb{\nonumber}
\def\ga{\left(}
\def\dr{\right)}
\def\aga{\left\{}
\def\adr{\right\}}

\def\rar{\rightarrow}
\def\lrar{\Longrightarrow}
\def\nnb{\nonumber}
\def\la{\langle}
\def\ra{\rangle}
\def\nin{\noindent}
\def\ba{\begin{array}}
\def\ea{\end{array}}

\def\als{\alpha_s}
\def\as{\ga\frac{\bar{\alpha_s}}{\pi}\dr}
\def\asr{\ga\frac{{\alpha_s}}{\pi}\dr}

\begin{document}
\topmargin -1.5cm
\oddsidemargin +0.2cm
\evensidemargin -1.0cm
\pagestyle{empty}
\begin{flushright}
PM/97-54\\ 
\end{flushright}
\vspace*{5mm}
\begin{center}
\section*{Running charm quark mass versus $M_D$ and $D_{(s)}\rar \mu\nu$}
\vspace*{1.5cm}
{\bf Stephan Narison}
\\
\vspace{0.3cm}
Laboratoire de Physique Math\'ematique et Th\'{e}orique\\
UM2, Place Eug\`ene Bataillon\\
34095 Montpellier Cedex 05, France\\
E-mail:
narison@lpm.univ-montp2.fr\\
\vspace*{2.5cm}
{\bf Abstract} \\ \end{center}
\vspace*{2mm}
\noindent
I study (for the first time) the dependence of $M_D$ and the leptonic decay constant$f_{D}$ on the
variation of the running charm quark mass $\bar m_c(\nu)$. I conclude that the present data 
on $f_{D_s}$ from $D_s\rar\mu\nu$ decay give a weaker constraint than $M_D$, where the latter
leads to the result:
$\bar m_c(M_c)=(1.08\pm 0.11)$ GeV,  to two-loop accuracy in the $\overline{MS}$ scheme, in agreement
with the value
$\bar m_c(M_c)=(1.23^{+0.04}_{-0.05})$ GeV extracted directly, within the same approximation,
from $M_{J/\psi}$. The agreement of the corresponding perturbative pole mass with the
one extracted directly from the data can indicate that the non-perturbative effects to the pole mass are
negligible. Inversely, injecting the average value of $\bar m_c(M_c)$ from $M_D$
and $M_{J/\psi}$ into the $\bar m_c$ behaviour of $f_{D}$, I obtain
$f_{D}\simeq (1.52\pm 0.16)f_\pi$, which combined with the sum rule prediction for
$f_{D_s}/f_D$, gives
$f_{D_s}\simeq (1.75\pm 0.18)f_\pi$ in good agreement with the data. 
The extension of the analysis to
the case of $f_B$ and $f_{D^*}$ is discussed. 
\noindent
\vspace*{7cm}
%\rule[.1in]{15.0cm}{.002in}   
\begin{flushleft}
PM/97-54 \\
December 1997
\end{flushleft}
\vfill\eject
\pagestyle{plain}
\setcounter{page}{1}
\section{Introduction}
One of the most important parameters of the standard model is the
quark masses. However, contrary to the leptons, where the physical mass
can be identified with the pole of the propagator, the quark masses are
difficult to define because of confinement. Some attempts have been
made in order to define the quark pole mass within perturbation theory,
where it has been shown to be IR-finite \cite{TARRACH} and independent
from the choice of the regularization and renormalization schemes used \cite{SNPOLE}.
More recently, it has been noticed, in the limit of a large number of flavours,
that the resummation of perturbative series can induce a non-perturbative term,
which can affect the truncated perturbative result, 
and can, then, limit the accuracy of
the pole mass determination \cite{BENEKE}.  However, a proper use of such a result 
needs a resummation, at the same
level of accuracy (which is not often the case), of the Green function of a given
process involving the pole mass, where some eventual cancellation
between the resummed terms of the two series may occur. One may bypass the
previous problems, by working, at a given order of
perturbative QCD, with the running quark masses, which are treated like coupling
constants of the QCD Lagrangian (see e.g. \cite{FNR}), and where some 
non-perturbative-like effect is expected to be absent. 
%\footnote{The observation from lattice
%calculations \cite{SOMMER} that the running mass follows its behaviour predicted by
%perturbation theory at a low scale may favour this expectation.}. 
A lot of efforts has been furnished in the
literature
\cite{PDG} for extracting directly from the data the running masses of the light
 and heavy quarks 
%\footnote{To my knowledge, the first determination of the
%light quark running mass has been done in \cite{BECCHI}, while the one of
% the running heavy quark masses is in
%\cite{SNM}.}, 
using the SVZ QCD spectral sum rules (QSSR) \cite{SVZ,SNB}. In this note, I
shall consider a direct extraction of the running charm quark mass
using the observed value of $M_D=1.865$ MeV and the new data on
$f_{D_s}$ from $D_s\rar\mu\nu$ decay, where its average value
is \cite{HEP96}:
\beq
f_{D_s}\simeq (1.92\pm 0.23)f_\pi.
\eeq
For convenience, we shall also use the fact that $r\equiv f_{D_s}/f_D$ is under good
control from different non-perturbative approaches. For definiteness and for
an internal self-consistency of the analysis, we shall use the QSSR value
\cite{SNFD}:
\beq
\frac{f_{D_s}}{f_D}=1.15\pm 0.04~
\lrar~ f_{D}\simeq (1.67\pm 0.24)f_\pi.
\eeq
We shall extend the analysis to the case of $f_B$ and $f_{D^*}$.
\section{The QCD spectral sum rules}
 We shall work with the pseudoscalar
two-point correlator: 
\beq
\psi_5(q^2) \equiv i \int d^4x ~e^{iqx} \
\la 0\vert {\cal T}
J_q(x)
J^\dagger _q(0) \vert 0 \ra ,
\eeq
built from the heavy-light quark current:
$
J_d(x)=(m_c+m_d)\bar c(i\gamma_5)d,
$
and which has the quantum numbers of the $D$ meson.
The corresponding Laplace transform sum rules are:
\beq
{\cal L}(\tau)
= \int_{t_\leq}^{\infty} {dt}~\mbox{e}^{-t\tau}
~\frac{1}{\pi}~\mbox{Im} \psi_5(t),~~~{\mbox {and}}~~~
{\cal R}(\tau) \equiv -\frac{d}{d\tau} \log {{\cal L}(\tau)},
\eeq
where $t_\leq$ is the hadronic threshold. The latter sum  rule,
 or its slight modification, is also useful, as it is equal to the 
resonance mass squared, in  
 the simple duality ansatz parametrization of the spectral function:
\beq
\frac{1}{\pi}\mbox{ Im}\psi_5(t)\simeq 2f^2_DM_D^4\delta(t-M^2_D)
 \ + \ 
 ``\mbox{QCD continuum}" \Theta (t-t_c),
\eeq
where the ``QCD continuum comes from the discontinuity of the QCD
diagrams, which is expected to give a good smearing of the
different radial excitations \footnote{At
the optimization scale, its effect is negligible, such that a more
involved parametrization is not necessary.}. The decay constant $f_D$ is
analogous to $f_\pi=93.3$ MeV; 
$t_c$ is the QCD continuum threshold, which is, like the 
sum rule variable $\tau$, an  a priori arbitrary 
parameter. In this
paper, we shall impose the
 $t_c$ and $\tau$ stability criteria for extracting our optimal
results \footnote{The corresponding $t_c$ value very roughly indicates
the position of the next radial excitations.}. 
The QCD expression of the correlator
is known to two-loop accuracy
(see e.g. \cite{SNB} and the explicit expressions given in \cite{SNFB}),
in terms  of the perturbative pole mass $M_c$, and including the non-perturbative
condensates of dimensions less than or equal to six
\footnote{We shall use the corrected coefficient of the quark-gluon mixed
condensate given in \cite{SOTTO}. This change affects only slightly the result. We shall
also skip the negligible contribution from the dimension six four-quark 
and three-gluon condensates. Notice that
there is some discrepancy on the value of the four-quark coefficient in the literature.}. The sum rule
reads:
\beq
{\cal L}(\tau)
= M^2_c\aga\int_{4M^2_c}^{\infty} {dt}~\mbox{e}^{-t\tau}~\frac{3}{8\pi^2} t(1-x)^2\Big{[}
1+\frac{4}{3}\asr f(x)\Big{]} +\Big{[} C_4\la O_4\ra +C_6\la
O_6\ra\tau\Big{]}~\mbox{e}^{-M^2_c\tau}\adr~,
\eeq
with:
\bea
x&\equiv& M^2_c/t,\nnb\\
f(x)&=&\frac{9}{4}+2\rm{Li}_2(x)+\log x \log (1-x)-\frac{3}{2}\log (1/x-1)\nnb\\
& & -\log (1-x)+ x\log (1/x-1)-(x/(1-x))\log x, \nnb\\
C_4\la O_4\ra&=&-M_c\la \bar dd\ra -\la \als G^2\ra/12\pi\nnb\\
C_6\la O_6\ra&=&\frac{M^3_c\tau}{2}
g\la\bar d\sigma_{\mu\nu}\frac{\lambda_a}{2}G_a^{\mu\nu}d\ra
%\nnb\\ &&-\ga\frac{8\pi}{27}\dr\ga 2-\frac{M^2_c\tau}{2}-\frac{(M^2_c\tau)^2}{6}\dr\rho\als \la \bar
%\psi\psi\ra^2~.
\eea 
It can be expressed in terms of the running mass $\bar{m}_c(\nu)$
\footnote{It is clear that, for the non-perturbative terms which are known to leading order
of perturbation theory, one can use either the running or the pole mass. However, we shall see
that the non-perturbative effects are not important in the analysis, such that this distinction
does not affect the result.}, through
the one-loop relation \cite{SNPOLE}:
\beq
M_c(\nu)=\bar{m}_c(\nu)\aga 1+\as\ga \frac{4}{3}+2\log{\frac{\nu}{M_c}}\dr
+...\adr .
\eeq
Throughout this paper we shall use the values of the parameters \cite{SNB,SNG}  given in Table 1.
We have used for the mixed condensate the parametrization:
\bea
g\la\bar d\sigma_{\mu\nu}\frac{\lambda_a}{2}G_a^{\mu\nu}d\ra&=&M^2_0\la\bar dd\ra.
%\nnb\\
%\la\bar d\Gamma_1\psi\bar d\Gamma_2\psi\ra &\simeq &
%{1}/({16N^2_c}){[} Tr{\Gamma_1}Tr{\Gamma_2-Tr{\Gamma_1\Gamma_2}}{]}
%\rho\la \bar
%\psi\psi\ra^2
\eea
%where $\rho\simeq 2\sim 3$ indicates the deviation from the vacuum saturation 
%assumption. 
We shall also use, for four active flavours \cite{BETHKE,PDG}:
\beq
\Lambda= (325\pm 100)~\mbox{MeV}.
\eeq
One can inspect that 
the dominant non-perturbative contribution  is due to the dimension-four $m_c\la \bar
dd\ra$ light quark condensate, while the other non-perturbative
effects remain a small correction
at the optimization scale, which corresponds to $\tau\simeq 0.6\sim 1$ GeV$^{-2}$ and $t_c
\simeq 6\sim 8$ GeV$^2$.\\
\section{Discussions and results }
\begin{figure}[H]
\begin{center}
\includegraphics[width=9cm]{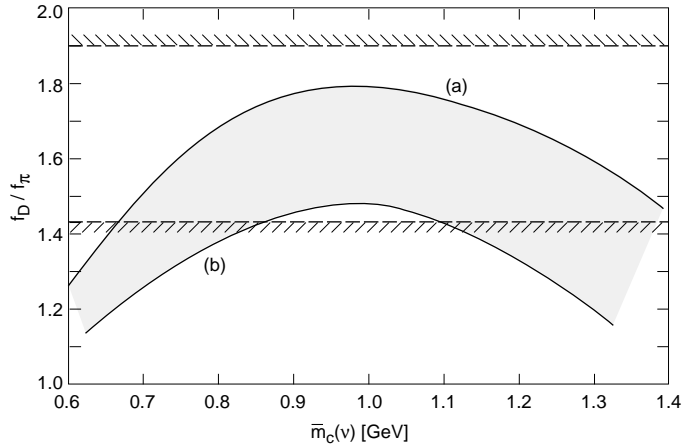}
\caption{Behaviour of $f_D$ versus $\bar m_c(\nu)$. The
horizontal band is the experimental domain of $f_D$. The theoretical band is limited by the
two curves: (a) $\Lambda$=0.425 GeV, $\nu$=1 GeV $\approx \tau^{-1/2}$, $t_c\geq 7$ GeV$^{2}$,
$\la \bar dd\ra^{1/3}$(1 GeV)=$-238$ MeV, 
$M^2_0=0.7$ GeV$^2$, 
$\la \alpha_s G^2\ra=0.06$ GeV$^2$
 and (b)
$\Lambda$=0.225 GeV, $\nu\approx M_c=1.42~\rm{GeV}$, $t_c=6$ GeV$^{2}$,
$\la \bar dd\ra^{1/3}$(1 GeV)=$-220$ MeV, 
$M^2_0=0.6$ GeV$^2$, 
$\la \alpha_s G^2\ra=0.08$ GeV$^2$.}
\end{center}
\end{figure}
\nin
Given the experimental value
on $M_D=1.865$ GeV, we present our results on $f_D$ from the first sum rule
for different values of the charm quark running mass evaluated at
$p^2=\nu^2$ in Fig. 1. The second sum rule gives the prediction on $M_D$ for
each   value of the charm mass (Fig. 2). The theoretical band is limited by the two extremal
values of the QCD parameters used in Table 1. Notice that the effect of the errors of the
different input parameters is much smaller in the ratio of sum rule ${\cal R}(\tau)$. The previous
analysis leads to the prediction from a two-loop calculation in the $\overline{MS}$ scheme:
\beq
\bar m_c(M_c)=(1.08\pm 0.11)~\mbox{GeV},
\eeq
where one should notice that, within the present accuracy
of the data on $f_{D_s}$, the result comes mainly from the one used to reproduce
the experimental value of $M_D$. 
\begin{figure}[H]
\begin{center}
\includegraphics[width=9cm]{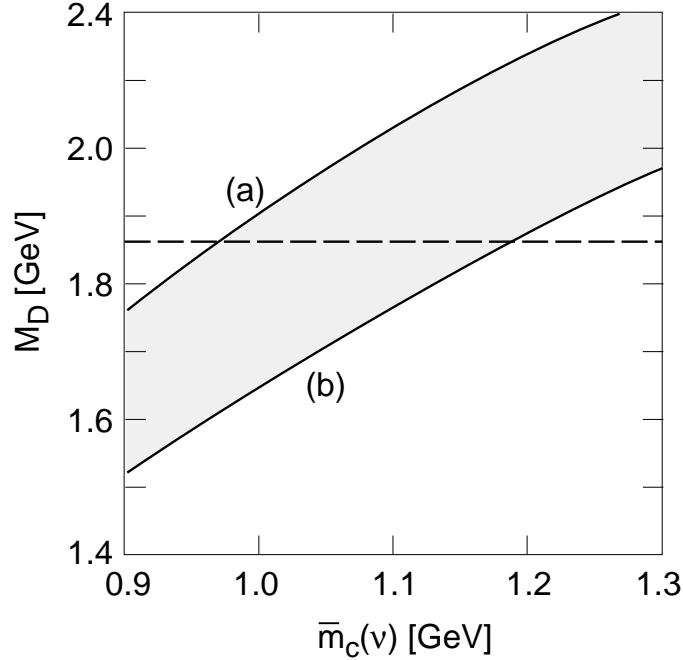}
\caption{Behaviour of $M_D$ versus $\bar m_c(\nu)$. The
horizontal band is the experimental value of $M_D$. The theoretical band is limited by the
two curves: (a) the same as (a) of Fig. 1 but $\nu$=1 GeV and (b)
the same as (b) of Fig. 1 but $\nu$=1.42 GeV.} 
\end{center}
\end{figure}
\nin
We have varied
the subtraction scale $\nu$ in the range from $\tau^{-1/2}\approx$ 1 GeV  to $M_c\approx$ 1.42 GeV. 
One should also notice, in Table 1, that the effect of $t_c$ on the result is relatively small from the
value $t_c\simeq 6$ GeV$^2$, where one starts to have a $\tau $ stability until $t_c\geq 7$ GeV$^2$,
where one has $t_c$ stability. The main sources of errors are due to $\Lambda$ and $\nu$.  
Within the errors, the present result is in good
agreement (though less accurate) with the value \cite{SNM}:
\beq
\bar m_c(M_c)=(1.23^{+0.04}_{-0.05})~\mbox{GeV}
\eeq
 obtained, within the same two-loop approximation,
from $M_{J/\psi}$. Inversely, we can use the combined value:
\beq
\bar m_c(M_c)=(1.20\pm 0.05)~\mbox{GeV}
\eeq
 from
$M_D$  and from $M_{J/\psi}$ systems
on the curve $f_D$ as function of $\bar m_c$ given in Fig. 1. Then, one can deduce:
\beq\label{fdrun}
f_{D}\simeq (1.52\pm 0.16)f_\pi,
\eeq
where, as can be seen in Table 1, the errors in this determination come mainly from the
perturbative parameters: $\Lambda$, $m_c$ and
the subtraction scale
$\nu$, and, to a lesser extent, from $t_c$ and the non-perturbative terms. 
We can compare this result with the previous value: 
\beq\label{fdpole}
f_{D}\simeq (1.35\pm 0.07)f_\pi,
\eeq
obtained by using the perturbative pole mass of the charm quark propagator \cite{SNFD,SNFB}. 
\begin{table*}[hbt]
\begin{center}
% space before first and after last column: 1.5pc
% space between columns: 3.0pc (twice the above)
\setlength{\tabcolsep}{1.2pc}
% -----------------------------------------------------
% adapted from TeX book, p. 241
%\newlength{\digitwidth} \settowidth{\digitwidth}{\rm 0}
%\catcode`?=\active \def?{\kern\digitwidth}
% -----------------------------------------------------
\caption{Different sources of errors in the estimate of $f_D$}
%\begin{tabular*}{\textwidth}{@{}l@{\extracolsep{\fill}}rrrrr}
\begin{tabular}{c c }
\hline 
%\hline
 & \\
Sources&$|\Delta (f_D/f_\pi)|$\\
&\\
\hline
%\hline
&\\
$\Lambda=(325\pm 100)$ MeV&0.12\\
$\nu=(1.20\pm 0.22)$ GeV&0.08\\
$\bar m_c(M_c)=(1.20\pm 0.05)$ GeV&0.05\\
$t_c=(6.5\pm 0.5)$ GeV$^2$&0.04\\
$\la \bar dd\ra^{1/3}$(1 GeV)=-$(229\pm 9)$ MeV&0.02\\
$M^2_0=(0.8\pm 0.1)$ GeV$^2$&0.02\\
$\la \alpha_s G^2\ra=(0.07\pm 0.01)$ GeV$^2$&0.01\\
%$\rho=2.5\pm 0.5$&0.005\\
&\\
Total& 0.16\\
&\\
\hline 
%\hline
\end{tabular}
\end{center}
\end{table*}
 The good
agreement between the two results in Eqs (\ref{fdrun}) and (\ref{fdpole}) within the errors may be an
indirect indication that the pole mass defined at a given order of perturbation theory, can provide a
good description of the physical process in this channel, and that the  eventual non-perturbative power
corrections induced by the resummation  of the perturbative series remain small corrections.
This observation can also be supported by the agreement of the value of the perturbative pole mass
obtained here and of the one from the $J/\psi$ systems, where both values have been obtained at two-loop
accuracy. A further support of this argument can also be provided by the agreement of the pole mass
deduced from Eq. (8) using the value of the running mass, with the one extracted directly in \cite{SNM}. 
Finally, using the previous value of the ratio
$ f_{D_s}/f_D$ given in Eq. (2), one obtains:
\beq
f_{D_s}\simeq (1.75\pm 0.18)f_\pi,
\eeq
which is in good agreement with the data quoted in Eq. (1). A natural improvement of the analysis done
in this paper is
a much more precise measurement of $f_{D_s}$ or/and an evaluation of the QCD two-point correlator to
three-loop accuracy. The two projects are expected to be feasible in the near future.
\section{Extension of the analysis to $f_B$ and $f_{D^*}$}
The extension of the previous analysis 
to the case of the $B$ meson is not very conclusive due to the strong dependence of the result on the
value of the subtraction scale $\nu$, which can vary, for the $B$ meson case, in a larger range. Instead,
using our previous observation on the agreement of the results on $f_D$ from the running and
perturbative pole masses, we also expect that the  result obtained from the pole mass of the
$b$ quark \cite{SNFD,SNFB} should also be reliable, which is:
\beq
f_{B}\simeq (1.49\pm 0.08)f_\pi.
\eeq
We have also extended the previous analysis to the case of the vector current having the quantum number
of the $D^*$. In this case, and working with the correlator having the same dimension as $\psi_5(q^2)$
\footnote{Notice that, in this channel, 
the correlator which has less power of $q^2$ than $\psi_5(q^2)$,  and which might present a $\tau$
stability, can have singularities and then should be used with a great care.},
we do not obtain a $\tau$ stability as the coefficients of the chiral condensates are opposite of the
pseudoscalar ones. Then, we conclude that, from this quantity, we cannot have a good
determination of $f_{D^*}$. 
\section*{Acknowledgements}
I thank,  for its hospitality, the CERN Theory Division, where this work has been finalized.
%\vfill\eject

\end{document}